\documentclass[12pt,dvips,usenames]{article}
\usepackage{color}
\usepackage{graphicx}
\usepackage{times}
\usepackage{amsmath}
\usepackage{amssymb}
\usepackage{subfigure}

\newcommand{\mbs}[1]{\boldsymbol{#1}}

 \def\bX{{\mbs{X}}}

\def\ba{{\mbs{a}}}

  \def\bl{{\mbs{l}}}
  
 \def\bq{{\mbs{q}}}

\def\cL{{{\cal L}}}
\def\cT{{{\cal T}}}
\def\cS{{{\cal S}}}

\def\cC{{{\cal C}}}

\title{An Analysis of the Quasicontinuum Method}
\date{\today}
\author{J.~Knap and M.~Ortiz \\
       \small\textsl{Graduate Aeronautical Labs}\\
       \small\textsl{California Institute of Technology}\\
       \small\textsl{Pasadena, CA 91125} }

\begin{document}

\maketitle

\subsection*{Abstract}
\label{Abstract}

{\small

The aim of this paper is to present a streamlined and fully
three-dimensional version of the quasicontinuum (QC) theory of
Tadmor {\it et al.} \cite{TadmorOrtizPhillips1996,
TadmorPhillipsOrtiz1996} and to analyze its accuracy and
convergence characteristics. Specifically, we assess the effect of
the summation rules on accuracy; we determine the rate of
convergence of the method in the presence of strong singularities,
such as point loads; and we assess the effect of the refinement
tolerance, which controls the rate at which new nodes are inserted
in the model, on the development of dislocation microstructures.

\vskip12pt
\noindent
{\it Keywords:} A. Quasicontinuum; B. Error analysis; C. Nanoindentation;
D. Atomistic models

\section{Introduction}
\label{sec:introduction}

The aim of this paper is to present a streamlined and fully
three-dimensional version of the quasicontinuum (QC) theory of
Tadmor {\it et al.} \cite{TadmorOrtizPhillips1996,
TadmorPhillipsOrtiz1996} and to analyze its accuracy and
convergence characteristics. The theory of the quasicontinuum
furnishes a computational scheme for seamlessly bridging the
atomistic and continuum realms. The chief objective of the theory
is to systematically coarsen an atomistic description by--and only
by--the judicious introduction of kinematic constraints. These
kinematic constraints are selected and designed so as to preserve
full atomistic resolution where required, e.~g., in the vicinity
of lattice defects, and to treat collectively large numbers of
atoms in regions where the deformation field varies slowly on the
scale of the lattice. Thus, in its purest form all input into the
theory concerning the behavior of the material is atomistic, and
all approximations are strictly kinematic in nature.

Different variants of the theory have been developed and
documented over a series of publications
\cite{TadmorOrtizPhillips1996, TadmorPhillipsOrtiz1996,
ShenoyMillerTadmor1998, MillerTadmorPhillips1998,
MillerOrtizPhillips1998, RodneyPhillips1999,
ShenoyMillerTadmor1999, TadmorMillerPhillips1999,
ShenoyPhillipsTadmor2000, SmithTadmorKaxiras2000}, where numerous
examples of application have also been presented. Here we endeavor
to emphasize the essential building blocks of the static theory,
to wit: i) the constrained minimization of the atomistic energy of
the solid; ii) the use of summation rules in order to compute the
effective equilibrium equations; iii) and the use of adaption
criteria in order to tailor the computational mesh to the
structure of the deformation field. In particular, we develop a
new class of summation rules based on sampling lattice functions
over clusters of atoms. In Section~\ref{sec:numer-exper}, we
present a detailed numerical analysis that probes various aspects
of the accuracy and performance of the method. Specifically, we
assess the effect of the cluster size and lumping procedure on
accuracy; we determine the rate of convergence of the method in
the presence of strong singularities, such as point loads; and we
assess the effect of the refinement tolerance, which controls the
rate at which new nodes are inserted in the model, on the
development of dislocation microstructures.

\section{The quasicontinuum method}
\label{sec:quas-meth}

We consider a reference configuration of a crystal such that its
$N$ atoms occupy a subset of a simple $d$-dimensional Bravais
lattice spanned by basis vectors $\{\ba_i, i= 1,\dots, d\}$. The
coordinates of the atoms in this reference configuration are:
\begin{equation}
  \label{eq:Bravais}
  \bX(\bl) = \sum_{i=1}^d l^i\mbs{a}_i,
  \quad \bl\in \cL \subset Z^d.
\end{equation}
Here, $\bl$ are the lattice coordinates which designate individual
atoms, $Z$ is the set of integer numbers and $d$ is the dimension
of space. From the definition, it follows that the set $\cL$ is
the collection of lattice sites occupied by atoms. 
The coordinates of the atoms in a deformed
configuration of the crystal are denoted $\{\bq(\bl),\
\bl\in~\cL\}$. For convenience, we shall collect all atomic
coordinates in an array $\bq$ and regard such array as an element
of the linear space $X \equiv R^{N d}$, the `configuration' space
of the crystal.

The energy of the crystal is assumed to be expressible as a
function $E(\bq)$, e.~g., through the use of empirical interatomic
potentials. In addition, the crystals may be subjected to applied
loads. We assume these loads to be conservative and to derive from
an external potential $\Phi^{\rm ext}(\bq)$. The total potential
energy is, therefore:
\begin{equation} \label{eq:PotentialEnergy}
\Phi(\bq) = E(\bq) + \Phi^{\rm ext}(\bq)
\end{equation}
The crystal may also be subject to displacement boundary
conditions over part of its boundary, e.~g., as a result of the
application of a rigid indentor. We wish to determine the stable
equilibrium configurations of the crystal. The problem is,
therefore, to determine all the \emph{local} minima of $\Phi(\bq)$
consistent with the displacement, or essential, boundary
conditions. We shall, somewhat equivocally, enunciate this problem
as:
\begin{equation}
  \label{eq:StaticProblem}
  \min_{\bq\in X} \Phi(\bq)
\end{equation}
We emphasize, however, that in general the aim is not simply to
determine the absolute minimizer of $\Phi(\bq)$, but rather the
much richer, and physically more relevant, set of metastable
configurations of the crystal.

The distinction arises from the fact that the energy function
$E(\bq)$ must be invariant under the class of affine deformations
which map the crystal lattice onto itself. This implies, for
instance, that $E(\bq)$ must be periodic under crystallographic
slip with period equal to the Burgers vector. Hence, the energy
function $E(\bq)$ is strongly nonconvex and the potential energy
$\Phi(\bq)$ in general has vast numbers of local minimizers, or
metastable configurations. 
Techniques for systematically exploring the full set of metastable
configurations, and indeed the entire configuration space of the
crystal, include statistical mechanics and path-integral methods.
An alternative approach is to load the crystal incrementally and
proceed by continuation. Thus, at every step in the solution
procedure the loads, or the prescribed displacements, are
incremented and the system is allowed to relax to a nearby stable
configuration. The relaxation process may be governed by inertia,
viscosity, kinetics, or some other rate-limiting mechanism; or may
be numerical in nature and owe to the use of iterative solvers
such as conjugate gradients. This latter approach is adopted in
all the numerical tests presented in this paper.

\subsection{Interpolation}
\label{sec:interpolation}

The essence of the theory of Tadmor {\it et al.}
\cite{TadmorOrtizPhillips1996, TadmorPhillipsOrtiz1996} is to replace
(\ref{eq:StaticProblem}) by a constrained minimization of
$\Phi(\bq)$ over a suitably chosen subspace $X_h$ of $X$. In order
to define $X_h$, we begin by selecting a reduced set
$\cL_h\subset\cL$ of $N_h < N$ `representative atoms'
(Fig.~\ref{fig:InterpolationFigureMesh}). The selection of the
representative atoms is based on the local variation of the fields
and will be addressed subsequently. In addition, we introduce a
triangulation $\cT_h$ of $\cL_h$. It bears emphasis that the
triangulation $\cT_h$ may be unstructured. In particular, ${\cal
L}_h$ need not define a Bravais lattice. The positions of the
remaining atoms are determined by piecewise linear interpolation
of the representative atom coordinates. We shall regard the
resulting coordinates $\{\bq_h(\bl),\bl\in\cL\}$ as belonging to a
linear space $X_h$ of dimension $N_h d$.

\begin{figure}
\begin{center}
\includegraphics[scale=1.0]{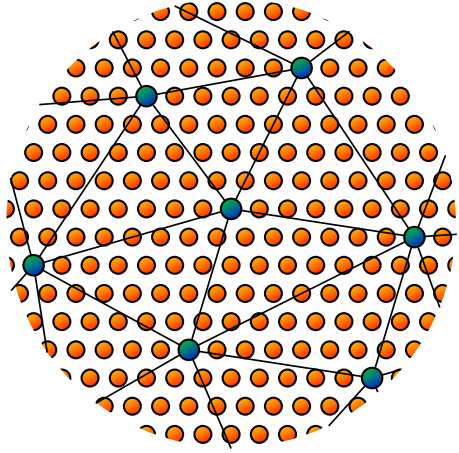}
\caption{Example of triangulation $\cT_h$ of the crystal.
\label{fig:InterpolationFigureMesh}}
\end{center}
\end{figure}

Let $\varphi_h(\bX | \bl_h)$, $\bl_h\in \cL_h$, be a collection of
shape functions for $\cT_h$. Thus, $\varphi_h(\bX | \bl_h)$ is
continuous and piecewise linear, its domain is restricted to the
simplices $K\in \cT_h$ incident to $\bX(\bl_h)$, and it vanishes
at all nodes of the triangulation except at $\bX(\bl_h)$, where it
takes the value $1$, i.~e.,
\begin{equation}
\varphi_h(\bX(\bl'_h) | \bl_h) = \delta(\bl'_h | \bl_h).
 \label{eq:ShapeFunctions}
\end{equation}
By construction,
\begin{equation}
\bq_h(\bl) = \sum_{\bl_h\in \cL_h} \varphi_h(\bl | \bl_h)
\bq_h(\bl_h), \label{eq:Interpolation}
\end{equation}
where we write
\begin{equation}
\varphi_h(\bl | \bl_h) = \varphi_h(\bX(\bl) | \bl_h).
\label{eq:ShortHand}
\end{equation}
Evidently, $\{\varphi_h(\bl | \bl_h),\ \bl_h\in \cL_h\}$
constitutes a basis for $X_h$ and the fields $\bq_h(\bl)$ are
entirely determined by their values $\bq_h(\bl_h)$ at the
representative atoms. In addition, the basis lattice functions are
required to satisfy the identity:
\begin{equation}
  \sum_{\bl_h\in\cL_h} \varphi_h(\bl | \bl_h) = 1
  \label{eq:PartitionUnity}
\end{equation}
i.~e., the basis lattice functions must define a partition of
unity over $\cL$. This requirement ensures that constant fields
are interpolated exactly by the basis lattice functions.

\subsection{Reduced equations}
\label{sec:reduced-equations}

The reduced counterpart of problem (\ref{eq:StaticProblem}) is now
\begin{equation}
  \min_{\bq_h\in X_h} \Phi(\bq_h).
  \label{eq:ReducedProblem}
\end{equation}
The minimizers of the reduced problem satisfy the reduced
equations of equilibrium
\begin{equation}
  \mbs{f}_h(\bl_h) = \sum_{\bl\in\cL}\mbs{f}(\bl | \bq_h)
  \varphi_h(\bl | \bl_h) = \mbs{0}.
\label{eq:ReducedEquilibrium}
\end{equation}
Here,
\begin{equation}
  \mbs{f}(\bq) = \Phi,_{\bq}(\bq)
  \label{eq:Forces}
\end{equation}
are the forces corresponding to $\bq$ and $\mbs{f}(\bl | \bq)$ is the
value of $\mbs{f}(\bq)$ at site $\bl$.  Thus, the reduced problem
entails the solution of the $N_h d$ equations
(\ref{eq:ReducedEquilibrium}) in the $N_h d$ unknowns $\bq(\bl_h)$,
$\bl_h\in \cL_h$.

\subsection{Node-based summation rules}
\label{sec:summation-rules}

The practicality of the method further hinges on the possibility
of avoiding the calculation of the full atomistic force array
$\mbs{f}$ and carrying out full lattice sums, as seemingly
required in (\ref{eq:ReducedEquilibrium}). As noted by Tadmor {\it
et al.} \cite{TadmorOrtizPhillips1996, TadmorPhillipsOrtiz1996}, this may be
accomplished by the introduction of summation rules similar to the
conventional quadrature rules of numerical integration.  The
problem is thus to approximate sums of the general form
\begin{equation}
  S = \sum_{\bl\in\cL} g(\bl),
  \label{eq:LatticeSum}
\end{equation}
where $g(\bl)$ is a lattice function. By analogy to numerical
quadrature rules, we begin by exploring summation rules of the
form
\begin{equation}
  S\approx \sum_{\bl\in\cS_h} n_h(\bl) g(\bl) \equiv S_h
  \label{eq:SummationRule}
\end{equation}
for some suitably chosen collection of summation points $\cS_h$
and weights $n_h(\bl)$, not necessarily integer.  Loosely
speaking, the weights $n_h(\bl)$ may be regarded as the number of
atoms represented by the site $\bl$. Proceeding as in the
development of numerical quadrature rules, the weights $n_h(\bl)$
may be determined by the requirement that the summation rule
(\ref{eq:SummationRule}) be exact for a restricted class of
lattice functions \cite{ShenoyMillerTadmor1999}.

The lowest-order summation rule is obtained by requiring that all
lattice functions in $X_h$, i.~e., all piecewise linear functions
supported by the triangulation $\cT_h$, be summed exactly. This is
tantamount to setting $\cS_h = \cL_h$ and requiring that the summation
rule (\ref{eq:SummationRule}) be exact for all shape functions
$\varphi_h(\bl | \bl_h)$, $\bl_h\in\cL_h$. This requirement gives,
explicitly,
\begin{equation}
  n_h(\bl_h) = \sum_{\bl\in\cL} \varphi_h(\bl | \bl_h), \qquad
  \bl_h\in\cL_h.
  \label{eq:SummationWeights}
\end{equation}
In fine regions of the triangulation, the sums on the right-hand
side of (\ref{eq:SummationWeights}) may be computed explicitly. By
contrast, in coarse regions of the mesh approaching the continuum
limit this explicit calculation becomes impractical. However, in
such regions the lattice sum (\ref{eq:SummationRule}) ostensibly
reduces to an integral and the corresponding weights are those of
conventional Lobatto quadrature \cite{hughes:1987}. In this limit,
each simplex $K \in \cT_h$ simply contributes $(N/V)|K|/(d+1)$
atoms to each of its $d+1$ nodes, where $N/V$ is the atom density
in the undeformed configuration of the crystal.

As a simple illustrative example, consider a monoatomic chain
discretized into `elements' each containing $L$ bonds. For an internal
node, the summation weight follows as
\begin{equation}
  n_h = 1 + 2 \sum_{l=1}^{L-1} \left(1 - \frac{l}{L}\right) = L,
  \label{eq:WeightsInterior}
\end{equation}
which coincides with the continuum limit.  By contrast, for an end
node the summation weight is
\begin{equation}
  n_h = 1 + \sum_{l=1}^{L-1} \left(1 - \frac{l}{L}\right) = \frac{L +
    1}{2}.
  \label{eq:WeightsBoundary}
\end{equation}
We note that $n_h$ tends to the continuum Lobatto limit of $L/2$ as
$L\to \infty$ and reduces to $n_h = 1$ in the full atomistic limit, as
required.

\subsection{Reduced equations based on summation rules}
\label{sec:reduc-equat-based}

The application of a summation rule to the evaluation of the
equilibrium equations (\ref{eq:ReducedEquilibrium}) leads to the
system of equations
\begin{equation}\label{eq:SummationEquilibrium}
  \mbs{f}_h(\bl_h) \approx
  \sum_{\bl\in\cS_h} n_h(\bl) \mbs{f}(\bl | \bl_h)
  \varphi_h(\bl | \bl_h) = {\bf 0}.
\end{equation}
As desired, the calculation of the effective forces in
(\ref{eq:SummationEquilibrium}) is of complexity ${\cal O}(N_h)$
provided that the number of sampling sites in $\cS_h$ is of order
$N_h$ and the atomic interactions are short-ranged.

The formulation of the method takes a somewhat unexpected turn at
this point in that the node-based summation rule discussed in the
foregoing suffers from a rank-deficiency problem. Thus, in the
linear case the node-based summation rule leads to rank-deficient
systems of equations. This is in analogy to the rank-deficiency of
finite-element stiffness matrices computed by Lobatto quadrature,
and manifests itself as the existence of so-called zero-energy
modes which pollute the displacement field or render the system of
equations singular altogether.

\begin{figure}
\begin{center}
\mbox{
  \includegraphics[scale=0.25]{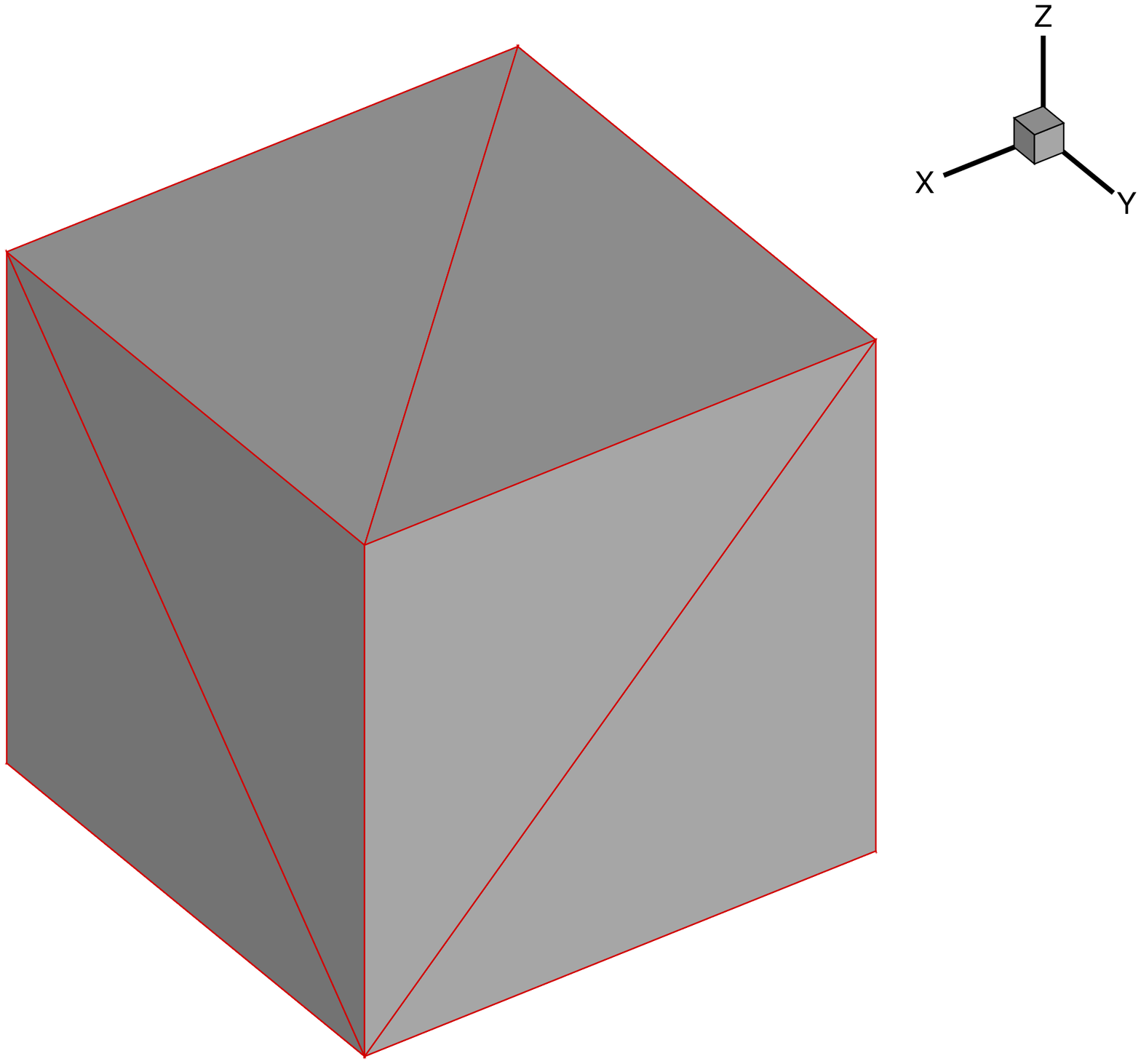}
  \includegraphics[scale=0.5]{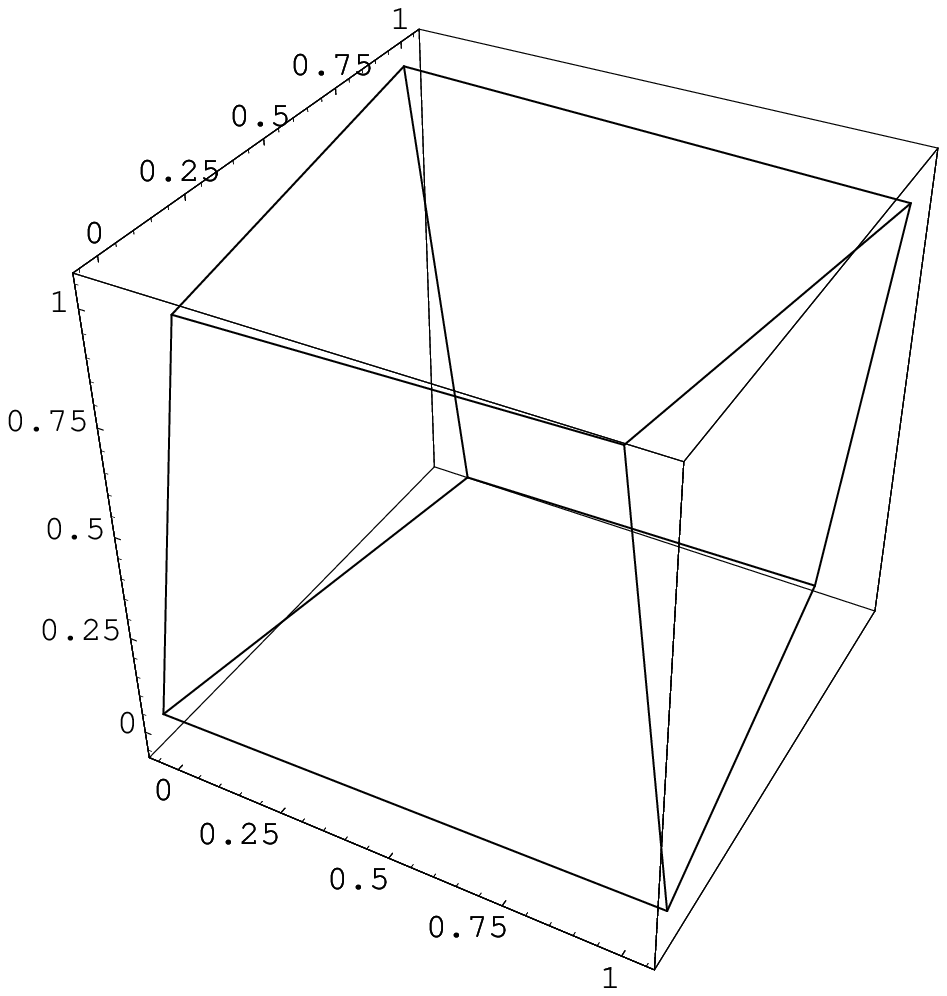}
  }
  \caption{a) Cubic Lennard-Jones crystal used in numerical
    stability tests. b) Zero-energy deformation mode}.
    \label{fig:StabilityCube}
\end{center}
\end{figure}

By way of illustration, we consider a Lennard-Jones crystal in the
form of a cube containing $32 \times 32 \times 32$ fcc unit cells.
In this example, eight representative atoms are introduced at the
corners of the cube, and the cube is triangulated as shown in
Fig.~\ref{fig:StabilityCube}a.  The stability of the crystal in
its reference configuration may be assessed by examining the
spectral properties of the reduced stiffness matrix. A direct
calculation of its eigenvalues reveals that seven of them are
identically equal to zero, which in turn denotes the presence of
one zero-energy deformation mode, Fig.~\ref{fig:StabilityCube}b.
However, the zero-energy deformation mode is absent when the
lattice sums are carried out in full, i.~e., when the test is
repeated with the equilibrium equations
(\ref{eq:SummationEquilibrium}) replaced by
(\ref{eq:ReducedEquilibrium}).

The numerical test just described shows that the node-based
summation rules are indeed rank-deficient, but it also suggests
that the reduced system is stable provided that a sufficient
number of sampling points is included in the summation rules. A
new set of summation rules which offers this added flexibility is
presented next.

\subsection{Cluster summation rules}
\label{sec:clust-based-summ}

\begin{figure}
\begin{center}
  \includegraphics[scale=1.0]{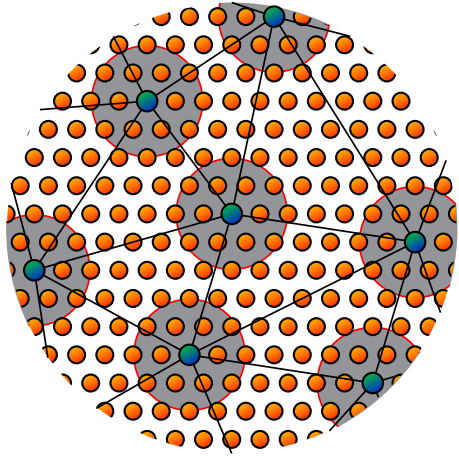}
  \caption{Clusters of atoms in triangulation $\cT_h$ of the crystal.
    \label{eq:ClustersFigure}}
\end{center}
\end{figure}

A class of summation rules which generalizes the node-based rule
may be obtained by sampling the lattice function over
neighborhoods of the representative atoms. We shall refer to these
neighborhoods as clusters, and the resulting summation rules as
cluster summation rules. Each cluster may be regarded as a
representative crystallite where the state and the behavior of the
crystal are sampled.

More specifically, let $\cC(\bl_h)=\{\bl : |\bX(\bl)-\bX(\bl_h)|
\leq r(\bl_h)\}$ be the cluster of lattice sites located within a
sphere of radius $r(\bl_h)$ centered on the representative atom
$\bl_h$ (Fig.~\ref{eq:ClustersFigure}). Assume for now that the
clusters thus defined do not overlap. The cluster summation rule
induced by the set of clusters $\cC(\bl_h), \bl_h \in \cL_h$ is
\begin{equation}
  S_h = \sum_{\bl_h \in \cL_h} n_h(\bl_h) S(\bl_h),
  \label{eq:ClusterSummationRule}
\end{equation}
where $S(\bl_h)$ denotes the sum over all atoms in the cluster
$\cC(\bl_h)$, i.~e.,
\begin{equation}
  S(\bl_h) = \sum_{\bl \in \cC(\bl_h)} g(\bl).
  \label{eq:ClusterSum}
\end{equation}
As before, the cluster weights $n_h(\bl_h), \bl_h \in \cL_h$ are
computed by requiring that the cluster summation rule
(\ref{eq:ClusterSummationRule}) be exact for all basis functions.

\begin{figure}
\begin{center}
  \includegraphics[scale=1.0]{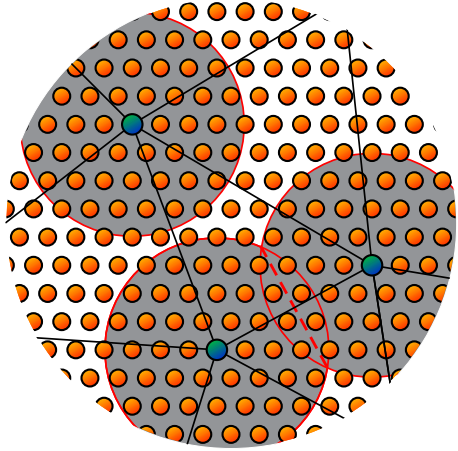}
  \caption{Triangulation $\cT_h$ with two overlapping clusters.
    The dashed line demarcates the boundary between the clusters.
    \label{eq:ClusterOverlapFigure}}
\end{center}
\end{figure}

It should be carefully noted that, as the triangulation size
approaches the atomic length scale, the spherical clusters defined
previously in general overlap, Fig.~\ref{eq:ClusterOverlapFigure}.
In this case, we truncate the clusters as follows. A site $\bl$
belongs to cluster $\cC(\bl_h)$ if its distance to $\bX(\bl_h)$ is
less than a prescribed value $r(\bl_h)$ and less than its distance
to any other representative atom. Thus, the cluster $\cC(\bl_h)$
is the intersection of the Voronoi cell containing $\bX(\bl_h)$
and sphere of radius $r(\bl_h)$ centered at $\bX(\bl_h)$.
Ambiguous cases corresponding to sites which are equidistant from
two representative atoms are resolved randomly. In the limit of
full atomistic resolution, each cluster contains exactly one atom
and the corresponding weight is $1$. Thus, the cluster summation
rules reduce seamlessly to the exact lattice sum when the fully
atomistic limit is attained.

\subsubsection{Summation-rule error analysis for monatomic chains}
\label{sec:error-analys-mono}

The accuracy of cluster summation rules can be analyzed simply in
the case of a one-dimensional monatomic chain. In this case, $\cL
\equiv Z$ is the set of all sites in the chain. Without loss of
generality, we take the coordinate of a site $l \in \cL$ in the
undeformed configuration of the chain to be $X(l) = l$. For
simplicity, we consider a uniform triangulation of $\cL$ of size
$h$. Thus, the coordinates of the representative atoms are $X(l_h)
= l_h h$, $l_h \in Z$, where $h$ is the simplex size. We consider
clusters of uniform radius $r < h/2$, Fig.~\ref{fig:1D_chain}.

\begin{figure}
\begin{center}
  \includegraphics[scale=1.0]{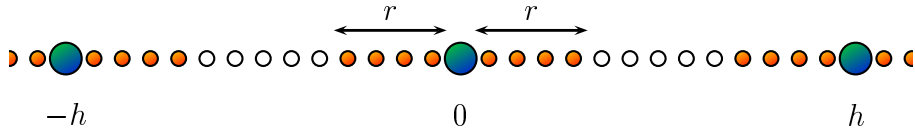}
  \caption{A part of the 1D monoatomic chain ($h=14$ and $r=4$).
    \label{fig:1D_chain}}
\end{center}
\end{figure}

The summation weights follow from the requirement that the basis
functions $\varphi_h(l|l_h)$ be summed exactly, with the result,
\begin{equation}
  n_h(l_h) = \frac{h}{1 + 2 r}, \qquad l_h \in Z.
  \label{eq:1D_chain_weight}
\end{equation}
Thus, for the infinite monatomic chain, $n_h(l_h)$ is simply the
ratio of the number of atoms in a simplex to the number of atoms
in a cluster.

In order to obtain the summation error to leading order, we
consider the sum of a quadratic polynomial supported on two
adjacent simplices in the reduced chain, namely,
\begin{equation}
    g(l) = \left\{
      \begin{array}{ll}
        l^2 & l \in [-h,h] \\
        0   & \mbox{otherwise}
      \end{array}
    \right.
  \label{eq:1D_chain_quadratic_function}
\end{equation}
The exact sum of $g(l)$ over all sites of the chain is
\begin{equation}
  S = \sum_{l \in \cL} g(l) = \sum_{l=-h}^{l = h} l^2 =
  \frac{h(1+h)(1+2 h)}{3}.
  \label{eq:1D_chain_exact_su}
\end{equation}
The calculation of the approximate sum $S_h$ requires three local
cluster sums $S_h(-h)$, $S_h(0)$ and $S_h(h)$
\begin{eqnarray}
  S_h(-h) & = & \sum_{l = -h}^{l = -h + r} l^2 =
  \frac{(1+r) (6 h^2 + r - 6h r + 2 r^2)}{6}, \\
  S_h(0) & = & \sum_{l = -r}^{l= r} l^2  =
  \frac{r (1+r)(1+2 r)}{3}, \\
  S_h(h) & = & \sum_{l = h-r}^{l=h} l^2  =
  \frac{(1+r)(6 h^2 + r - 6 h r + 2 r^2)}{6}.
  \label{eq:1D_chain_local_sums}
\end{eqnarray}
The use of the cluster summation rules leads to the approximate
lattice sum
\begin{equation}
  S_h = n_h [S_h(-h) + S_h(0) + S_h(h)] =
  \frac{2 h (1+r)}{3 (1+ 2 r)}
        [2 r^2 + r (1-3 h) + 3 h^2].
  \label{eq:1D_chain_approx_sum}
\end{equation}
The summation error is $|S_h - S|$. An upper bound on the error is
obtained by setting $r=0$. Then, for $h \gg 1$ we have $|S_h -
S|_{r=0} \sim (4/3) h^3$, which shows that the summation rule is
of third order.

The effect of the cluster size on the accuracy of the summation
rule may be understood as follows. For simplicity, we consider the
case of $h \gg 1$, whereupon (\ref{eq:1D_chain_exact_su})
simplifies to
\begin{equation}
    S \sim \frac{2}{3} h^3
    \label{eq:1D_chain_exact_su2}
\end{equation}
and (\ref{eq:1D_chain_approx_sum}) to
\begin{equation}
  S_h =
  \frac{2 (1+r)}{3 (1+ 2 r)} (2 \xi^2 - 3 \xi + 3 ) h^3
  \label{eq:1D_chain_approx_sum2}
\end{equation}
where we write $\xi = r/h$. In the regime of $\xi \ll 1$,
(\ref{eq:1D_chain_approx_sum2}) simplifies further and the
corresponding relative error behaves as
\begin{equation}
    \frac{|S_h - S|}{|S|} \sim \frac{2 + r}{1 + 2 r}
  \label{eq:SummationError1}
\end{equation}
which is a decreasing function of $r$. Thus, as expected,
increasing $r$ from 0 increases the accuracy of the summation
rule. In the regime of $r \gg 1$ and $r \sim h$, the relative
error behaves as
\begin{equation}
   \frac{|S_h - S|}{|S|} \sim  \frac{1}{2} (2 \xi - 1) (\xi - 1)
  \label{eq:1D_chain_relative_error}
\end{equation}
which vanishes for $\xi = 1/2$, or $r = h/2$. Thus, the exact sum
is obtained when the clusters encompass the entire lattice, as
required.

\subsection{Reduced equations based on cluster summation rules}
\label{sec:reduc-equat-based-1}

The application of the cluster based summation rule to the reduced
equilibrium equations (\ref{eq:ReducedEquilibrium}) yields
\begin{equation}
  \mbs{f}_h(\bl_h) \approx \sum_{\bl'_h \in \cS_h} n_h(\bl'_h)
               \left[ \sum_{\bl \in \cC(\bl'_h)}  \mbs{f}(\bl)
               \varphi_h(\bl | \bl_h)\right] = \mbs{0}.
   \label{eq:ClusterBasedEquilibrium}
\end{equation}

The computational complexity of (\ref{eq:ClusterBasedEquilibrium})
is ${\cal O}(N_h N_r)$, where $N_r$ is the number of lattice sites
in a cluster of radius $r$. For $N_r > 1$, the computational
effort is in excess of that corresponding to the node-based
summation rules. Thus, the optimal value of the cluster size
$r(\bl_h)$ is subject two opposing requirements. On one hand, the
cluster size has to contain a sufficient number of atoms to ensure
the appropriate accuracy and stability of the summation rule. 
On the other, however, such number should remain small for the 
method to retain its computational efficiency. The
effect of the cluster size on the accuracy is investigated
subsequently by way of numerical testing.

\subsection{Adaptive selection of representative atoms}
\label{sec:adapt-select-repr}

The third key component of the quasicontinuum method is the use of
mesh adaption in order to tailor the computational mesh to the
structure of the deformation field. Ideally, the adaptivity itself
should be driven by the energetics of the system, i.~e., the mesh
should return the least possible potential energy for a fixed
number of representative atoms. However, in order to optimize the
mesh in this manner it is necessary to know the relation between
the energy, or suitable energy bounds thereof, and the mesh size.
Unfortunately, unlike the continuum case, an approximation theory
for discrete systems such as considered here appears to be
entirely lacking at present, and such energy bounds are presently
unavailable.

Under these conditions, it becomes necessary to resort to
empirical adaption indicators. In calculations we adopt as
adaption indicator $\epsilon(K)$ for simplex $K$ the variation of
the displacement field over $K$ modulo rotations. This variation
should be well-defined since the displacement field of the crystal
may reasonably be expected to be of bounded variation.
Specifically, we measure the variation of the displacement field
as
\begin{equation}
\epsilon(K) \equiv \sqrt{|II_E(K)|} h(K)
\end{equation}
where $II_E(K)$ denotes the second invariant of the Lagrangian
strain tensor in simplex $K$, and $h(K)$ is the size of $K$ (cf
\cite{TadmorOrtizPhillips1996,TadmorPhillipsOrtiz1996}). It follows from its definition
that $\epsilon(K)$ is invariant under rotations. The element $K$
is deemed acceptable if
\begin{equation} \label{eq:error_strain2}
\frac{\epsilon(K)}{b} < TOL
\end{equation}
for some prescribed tolerance $TOL < 1$, and is targeted for
refinement otherwise. In (\ref{eq:error_strain2}) $b$ denotes the
magnitude of the smallest Burgers vector of the crystal.

In our implementation of the method, the refinement of the
simplices which violate (\ref{eq:error_strain2}) is accomplished
by the application of subdivision rules. In particular, we choose
to bisect the longest edge of the simplices which are targeted for
refinement. The new representative atoms are inserted in the
lattice site nearest to the geometrical midpoint of the longest
edge. In order to preserve the quality of the computational mesh,
we apply standard mesh-improvement operations such as edge-face or
octahedral swapping \cite{joe:1989, joe:1995, freitag:1996} and
smoothing \cite{shephard:1991}. Each new mesh is equilibrated and
the remeshing criterion (\ref{eq:error_strain2}) is re-evaluated
for the new solution. The process is repeated until the mesh
remains unchanged.


The significance of (\ref{eq:error_strain2}) becomes apparent by
considering a processes of crystallographic slip across $K$. To
this end, imagine that a pair of representative atoms in $K$
undergo a relative sliding displacement of magnitude $b$ across a
slip plane of the crystal. For this deformation one has
\begin{equation}
  \label{eq:error_strain1}
  \epsilon(K) = \frac{b}{h(K)}
\end{equation}
It follows, therefore, that the adaption criterion
(\ref{eq:error_strain2}) ensures that the mesh size $h(K)< b$
under the conditions just described. Thus, the adaption criterion
is designed so that full atomistic resolution is attained when a
simplex slips by a full Burgers vector. Interestingly, we have
found that the dislocation patterns predicted in calculations are
sensitive to the choice of tolerance, and generally $TOL$ must be
chosen much smaller than $1$ in order not to inhibit dislocation
nucleation.

\section{Details of the computer implementation}
\label{sec:deta-comp-impl}

In order for the quasicontinuum method to perform optimally, some
implementational issues require careful attention. Several
potential performance bottlenecks are discussed in this section.

\subsection{Site-element mapping}
\label{sec:site-element-mapping}

The numerical solution of the quasicontinuum equilibrium equations
(\ref{eq:SummationEquilibrium}), requires repeated computation of
atomic coordinates $\bq_h(\bl),\ \bl \in \cL$ which  depend,
through the interpolation constraints (\ref{eq:Interpolation}), on
the coordinates of representative atoms $\bq_h(\bl_h),\ \bl_h \in
\cL_h$. It is therefore important to devise an efficient algorithm
for locating the simplex $K \in \cT_h$ which contains a given site
$\bl$ of the crystal.

The simplex $K \in \cT_h$ containing site $\bl \in \cL$ may be
found from elementary geometry and an exhaustive search over
$\cT_h$. Such method carries, however, substantial computational
cost, due to a large number of required floating-point operations.
Our approach is based on the fact that once the pair $\{\bl, K\}$
has been established, it remains valid for the lifespan of
$\cT_h$.  Thus, the following two stage strategy may be adopted:
\begin{enumerate}
\item A general search routine, which for a given lattice site $\bl
  \in \cL$ and triangulation $\cT_h$ returns $K \in \cT_h$ containing
  $\bl$.
\item A look-up table (cache) that stores already associated pairs
  $\{\bl, K\}$.
\end{enumerate}
At first, the look-up table contains no $\{\bl, K\}$ pairs and
most of the inquiries results in the execution of the general
search routine. As the force calculation proceeds, pairs are
continuously inserted into the table, and fewer calls to the
general routine are needed. At the end of the first force
calculation, most of the pairs can be located in the cache, where
they remain as long as $\cT_h$ stays unchanged.

Our implementation of the look-up table is based on hashing, which
is the method of referencing records in a table by doing
arithmetic transformations of keys into table addresses. Any
hashing algorithm requires two design decisions to be made:
\begin{enumerate}
\item A hash function $h(k)$ taking a key $k$ as its argument and
  returning an index into a table must be chosen.
\item One must establish a strategy for dealing with cases of two
  distinct keys $k_1$ and $k_2$ for which the resulting values of the
  hash function are the same, i.~e. $h(k_1) = h(k_2)$ (collision
  resolution).
\end{enumerate}

\noindent In his monograph, Knuth \cite{knuth97} provides an
excellent description of hashing, including some possible choices
of hash functions. For a simple three-dimensional Bravais lattice
(\ref{eq:Bravais}), the lattice coordinates of a site $\bl =
\{l^1, l^2, l^3\} \in Z^3$ may be considered as a key. The hash
function acting on this key produces an index into a table of
simplices where the $K$, containing $\bl$, can be found. In cases
in which a key contains more than one word (each of $l^i$ can be
considered as a separate word) Knuth \cite{knuth97} suggests using
the following hash function
\begin{equation}
  \label{eq:Hash_function}
  h(\bl) = [h_1(l^1) + h_2(l^2) + h_3(l^3) ] \mod M.
\end{equation}
where $h_i(k)$ denotes a hash function for $l^i$ and M is an
integer parameter. The function $a \mod p$ computes the remainder
of $a/p$. In general, the optimal choice of $h_i$ depends on
$\cT_h$ and cannot easily be established. We have found, however,
that the particular hash function
\begin{equation}
  \label{eq:Hash_function_i}
  h_i(l^i) = \mbox{lshift}(l^i, 2^{i-1}), \qquad i=1,2,3.
\end{equation}
works well for a wide range of triangulations. The meaning of
$\mbox{lshift}(l, m)$ is simply ``shift all bits in $l$ left $m$
positions''. The operations involved in the calculation of $h_i$
are elementary, and the computational cost of each such
calculation is small. The integer parameter $M$ in
(\ref{eq:Hash_function}) may be chosen arbitrarily, but Knuth
\cite{knuth97} points out that, when it is taken to be some power
of $2$, i.~e., $M=2^p$, $a \mod M$ is equivalent to masking the
low $p$-bits from $a$ (on most of currently available computer
hardware).  The application of (\ref{eq:Hash_function}) to a
lattice site $\bl$ results in an index into the hash table from
the interval $[0, 2^p)$. Accordingly, the size of the table is
limited to $2^p$, which renders collisions likely. The collisions
are handled with the help of additional short tables which store
$\{\bl,K\}$ pairs for all $\bl$ such that the $h(\bl)$ produces
the same value. Therefore, the search through the cache for a
simplex $K$ containing site $\bl$ becomes essentially a two stage
process, in which the application of the hash function is followed
by a linear search. Since, the number of lattice sites having the
same index is small, and the additional cost due to the linear
search phase is not significant.

\subsection{Computation of cluster weights}
\label{sec:comp-clust-weights}

As explained earlier, the cluster weights $n_h(\bl_h),\ \bl \in
\cL_h$ are obtained requiring that the shape functions are summed
exactly by the cluster summation rule
(\ref{eq:ClusterSummationRule}). The application of the summation
rule to shape functions $\varphi_h(\bl|\bl_h)$ leads to the system
of $N_h$ linear algebraic equations
\begin{equation}
  \label{eq:Weight_system}
    \sum_{\bl'_h \in \cL_h} A(\bl_h|\bl'_h) \, n(\bl'_h) = b(\bl_h),
    \qquad \bl_h \in \cL_h,
\end{equation}
to be solved for $n_h(\bl_h),\ \bl \in \cL_h$. Here the matrix
$A(\bl_h|\bl'_h),\ \bl'_h, \bl_h \in \cL_h$ is
\begin{equation}
  \label{eq:Weight_system_matrix}
  A(\bl_h|\bl'_h) = \sum_{\bl' \in \cC(\bl'_h)} \varphi_h(\bl|\bl_h).
\end{equation}
and $\cC(\bl'_h)$ denotes the cluster of lattice sites centered at
$\bl'_h$. The array $b(\bl_h),\ \bl_h$ is simply
\begin{equation}
  \label{eq:Weight_system_rhs}
  b(\bl_h) = \sum_{\bl \in \cL} \varphi_h(\bl|\bl_h).
\end{equation}
The calculation of the array $b$ requires the evaluation of the
shape functions at all lattice sites within a crystal. While this
may be regarded as acceptable for small crystals, it becomes
prohibitively expensive when applied to large samples. However, as
we have already shown, in the continuum limit summation may be
replaced by integration, in which case simplex $K$ contributes
$N(K)/(d+1)$ to each of its vertices. $N(K)= (N/V) |K|$ is the
approximate number of sites within $K$ and $N/V$ is the atom
density in the undeformed configuration of the crystal.
Specifically, one may introduce a cutoff $N_c$ and restrict the
direct calculation of the sum in (\ref{eq:Weight_system_rhs}) to
simplices $K$ for which $N(K) \leq N_c$.

It is apparent from (\ref{eq:Weight_system_matrix}) that $A$ has a
sparse structure, which suggests the use of specialized solvers
for sparse linear systems. However, a simpler alternative route is
to resort to lumping in order to replace $A$ by a diagonal matrix
(see, e.~g., \cite{hughes:1987}). A lumping technique which is
widely used in finite elements is the row-sum technique, which
gives the diagonal entries of the lumped matrix as
\begin{equation}
  \label{eq:Weight_system_matrix_diag}
  A(\bl_h|\bl_h) = \sum_{\bl'_h \in \cL_h} \sum_{\bl' \in \cC(\bl'_h)}
    \varphi(\bl|\bl_h), \qquad \bl_h \in \cL_h,
\end{equation}
with all other entries in $A$ set to zero. Once the matrix $A$ is
lumped, the solution of (\ref{eq:Weight_system_matrix}) is trivial
and gives
\begin{equation}
  \label{eq:Weight_system_diag_solve}
  n(\bl_h) = \frac{b(\bl_h)}{A(\bl_h|\bl_h)}, \qquad \bl_h \in \cL_h.
\end{equation}
In the atomistic limit, the values of cluster weights computed
from (\ref{eq:Weight_system_diag_solve}) become identical with
obtained exactly.

\section{Numerical tests}
\label{sec:numer-exper}

The accuracy of the quasicontinuum method is largely determined by
three factors:
\begin{itemize}
\item The value of the cluster-size parameter $r$, which controls
    the accuracy of the cluster summation rules.
\item The approximations introduced in the computation of the
  cluster weights, namely: the approximate calculation of vector
  $b$, controlled by the cuttoff parameter $N_c$; and the lumping
  procedure for constructing the diagonal matrix $A$.
\item The value of parameter $TOL$ in (\ref{eq:error_strain2}),
    which controls the process of representative atom insertion.
\end{itemize}

In this section we present an analysis of the influence of each of
these factors on the accuracy of the quasicontinuum method. We
measure the quality of the solution in energy terms. Specifically,
we identify the error in the approximate solution $\bq_h$ with
\begin{equation}\label{eq:Energy_error}
e = \Phi(\bq_h) - \Phi(\bq)
\end{equation}
where $\bq$ is the solution of the full system obtained, e.~g., by
relaxing $\bq_h$. Since the potential energy decreases during this
relaxation, the energy error is greater or equal to zero. For a
well-supported harmonic crystal the energy error defined in
(\ref{eq:Energy_error}) defines a proper norm of the error lattice
function $\bq_h - \bq$.

\subsection{Test problem definition}

\begin{figure}
\begin{center}
  \includegraphics[scale=1.0]{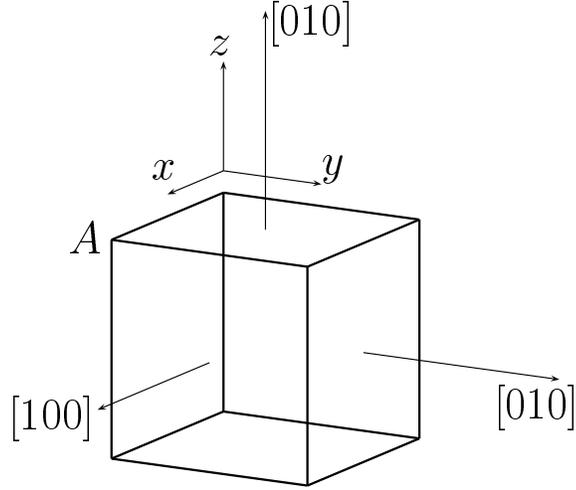}
  \caption{Crystallographic orientation of test sample used in the
    simulations of nanoindentation.
    \label{fig:Convergence_sample}}
\end{center}
\end{figure}

\begin{figure}
\begin{center}
  \includegraphics[scale=0.5]{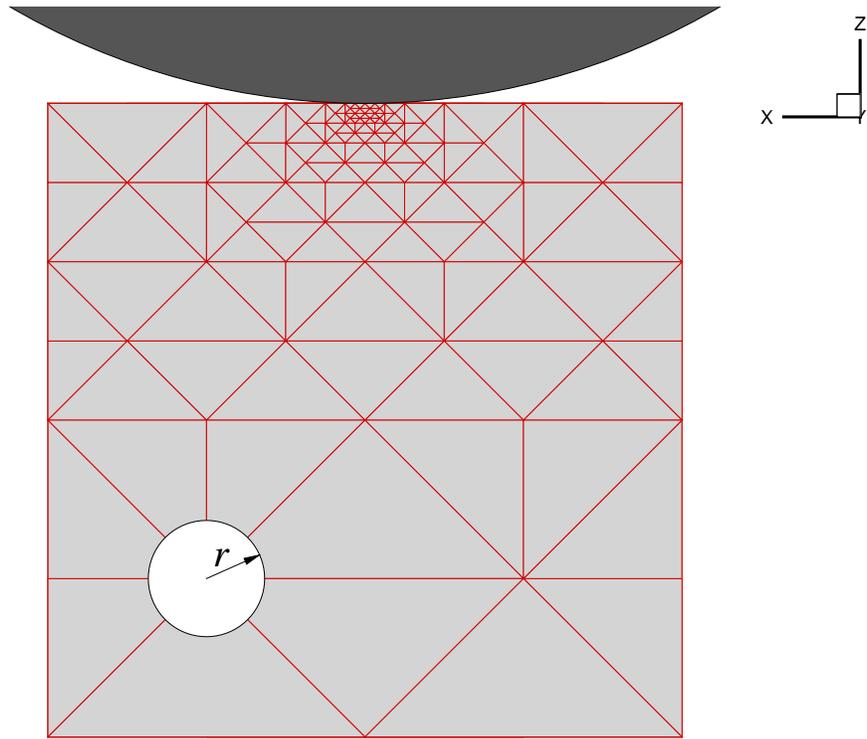}
  \caption{Cross section through the quasicontinuum test sample
    used in the nanoindentation simulations. A part of the indentor is
    also shown. The white circle represents a cluster of radius
    $r=\sqrt{69} [\sigma]$ (corresponding to sixty-four shell of
    neighbors).\label{fig:Initial_mesh}}
\end{center}
\end{figure}

We take \emph{nanoindentation} as a convenient test problem for
assessing the performance of the method. A salient feature of
nanoindentation is the presence of a highly nonuniform state of
deformation resulting in a sharp mesh gradation away from the
indentor. This feature effectively tests the adaptivity of the
method.

The test sample used in simulations is a cube of an fcc
nearest-neighbor Lennard-Jones crystal containing $64 \times 64
\times 64$ fcc unit cells, or a total of $1\,073\,345$ atoms. The
size of the cube is limited by the need to compute the solution of
the full problem by direct atomistic simulation in order to
evaluate the error. The Lennard-Jones potential is
\cite{lennard-jones39:_critic, lennard-jones39a:_critic}
\begin{equation}
\phi(r) = 4 \varepsilon \left[ \left(\frac{\sigma}{r}\right)^{12}
- 2 \left(\frac{\sigma}{r}\right)^6 \right],
\label{eq:Lennard_Jones}
\end{equation}
where $\sigma$ and $\varepsilon$ are parameters which set the
length and energy scales, respectively. The choice of the
Lennard-Jones potential is motivated by the desire to eliminate
surface effects which would otherwise compound the interpretation
of results.

The surfaces of the sample are aligned with the cube directions,
Fig.~\ref{fig:Convergence_sample}. Atoms in the bottom surface are
constrained to remain at their initial positions throughout the
test, whereas in all side surfaces the atoms are allowed to move
in the $z$-direction only.  No displacement constraints are
introduced on the top surface of the cube.

In calculations we use a model of a spherical indentor proposed by
Kelchner {\it et al.} \cite{kelchner98:_disloc}. In this model,
the indentor is regarded as an additional external potential
$\Phi^{\rm ext}$ interacting with atoms in the substrate. The
potential has the particular functional form
\begin{equation}
  \Phi^{\rm ext}(r) = A H(R-r) (R-r)^3,
  \label{eq:indentor_potential}
\end{equation}
where $R$ is the radius of the indentor, $r$ denotes distance
between a site and the center of the indentor, $A$ is a force
constant and $H(r)$ is the step function. In calculations we adopt
the following values of the parameters: $R=100\ [\sigma]$ and
$A=2000\ [\varepsilon/\sigma^3]$.

The initial triangulation of the cube is specifically tailored to
the nanoindentation geometry, Fig.~\ref{fig:Initial_mesh}. Thus,
in the region of the crystal located directly underneath the
indentor, full atomistic resolution is introduced from the outset.
Away from this region, the triangulation becomes gradually
coarser. The resulting number of representative atoms in the
initial mesh is $888$, a significant reduction from the total
number of atoms ($1\,073\,345$) in the sample. A cross section of
the initial mesh through the center of the cube with a plane
$y=const$ is show in Fig.~\ref{fig:Initial_mesh}. All solutions
are computed using a nonlinear version of the Conjugate Gradient
method.

\subsection{Effect of the cluster size}

\begin{figure}
\begin{center}
  \includegraphics[scale=0.5]{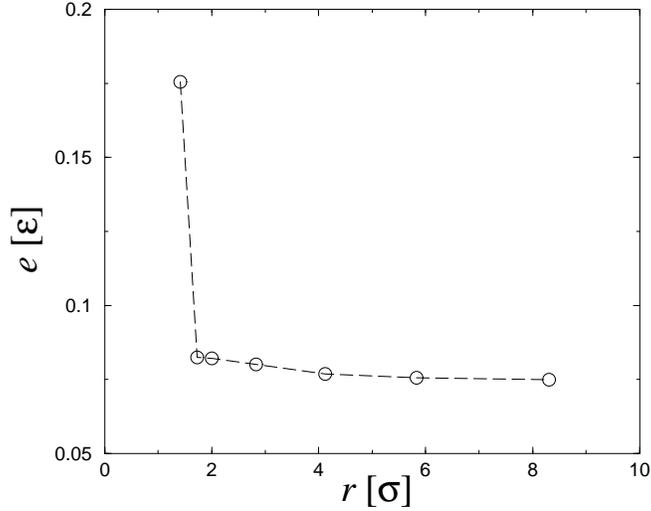}
  \caption{The error $e$ as a function of the cluster size
    $r$. \label{fig:Cluster_size_plot}}
\end{center}
\end{figure}

The cluster radius $r$ controls the accuracy of the cluster
summation rules. In order to isolate the effect of the cluster
size from other factors, in the present tests the summation
weights are computed exactly and the mesh is kept fixed. The
indentor is pushed through a distance $0.1\ [\sigma]$ into the
crystal along the z-axis. This value ensures that the material
immediately under the indentor becomes highly deformed and is
pushed well into the nonlinear regime. However, the induced
deformation is not sufficient for any defects to arise. This
greatly simplifies the interpretation of results and the error
analysis, since under the stated conditions the full atomistic
solution $\bq$ to compare with is well-defined.

The computed dependence of the error $e$ on the cluster size $r$
is shown in Fig.~\ref{fig:Cluster_size_plot}. It is evident from
this plot that the inclusion of a single shell of neighbors in the
clusters suffices to eliminate the rank-deficiency of the
node-based summation rule. As suggested by our previous analysis,
the error is greatest for the smallest cluster size and decreases
rapidly as the cluster size increases. For $r$ above $\sqrt{8}\
[\sigma]$, corresponding to the radius of the eighth shell of
neighbors, the solution becomes insensitive to the cluster size.
Thus, the use of relative small clusters results in high accuracy
while preserving the computational efficiency of the method.

\subsection{Effect of the lumping procedure}
\label{sec:infl-lump-proc}

\begin{figure}
\begin{center}
  \includegraphics[scale=0.5]{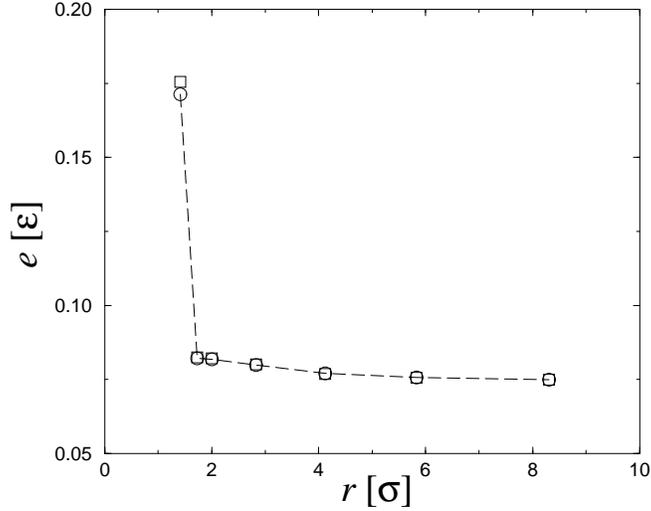}
  \caption{The error $e$ as a function of the cluster size
    $r$. Squares and circles correspond to the
    approximate and exact calculation of cluster weights,
    respectively.}
           \label{fig:Cluster_size_plot_lumped}
\end{center}
\end{figure}

The error in the calculation of the cluster weights originates
from two sources: the approximate computation of the array $b$ in
(\ref{eq:Weight_system_rhs}), controlled by the value of parameter
$N_c$; and the lumping of matrix $A$ in
(\ref{eq:Weight_system_matrix}). In order to appraise this error,
we repeat the calculations described in the preceding section with
the  summation weights computed approximately. The cutoff $N_c$ is
set to $2000$, i.~e. the contributions to the array $b$ from
simplices which contain fewer than $2000$ atoms are computed
explicitly. The dependence of the error $e$ on the cluster size
parameter $r$ for both the approximate (squares) and exact
(circles) weights is shown in
Fig.~\ref{fig:Cluster_size_plot_lumped}. Small differences do
arise for very small clusters, and virtually disappear for large
clusters. In view of these results we may conclude that the use of
approximate summation weights does not result in a significant
loss of accuracy.

\subsection{Convergence properties of the quasicontinuum method}
\label{sec:conv-prop-quasi}

\begin{figure}
\begin{center}
  \includegraphics[scale=0.27]{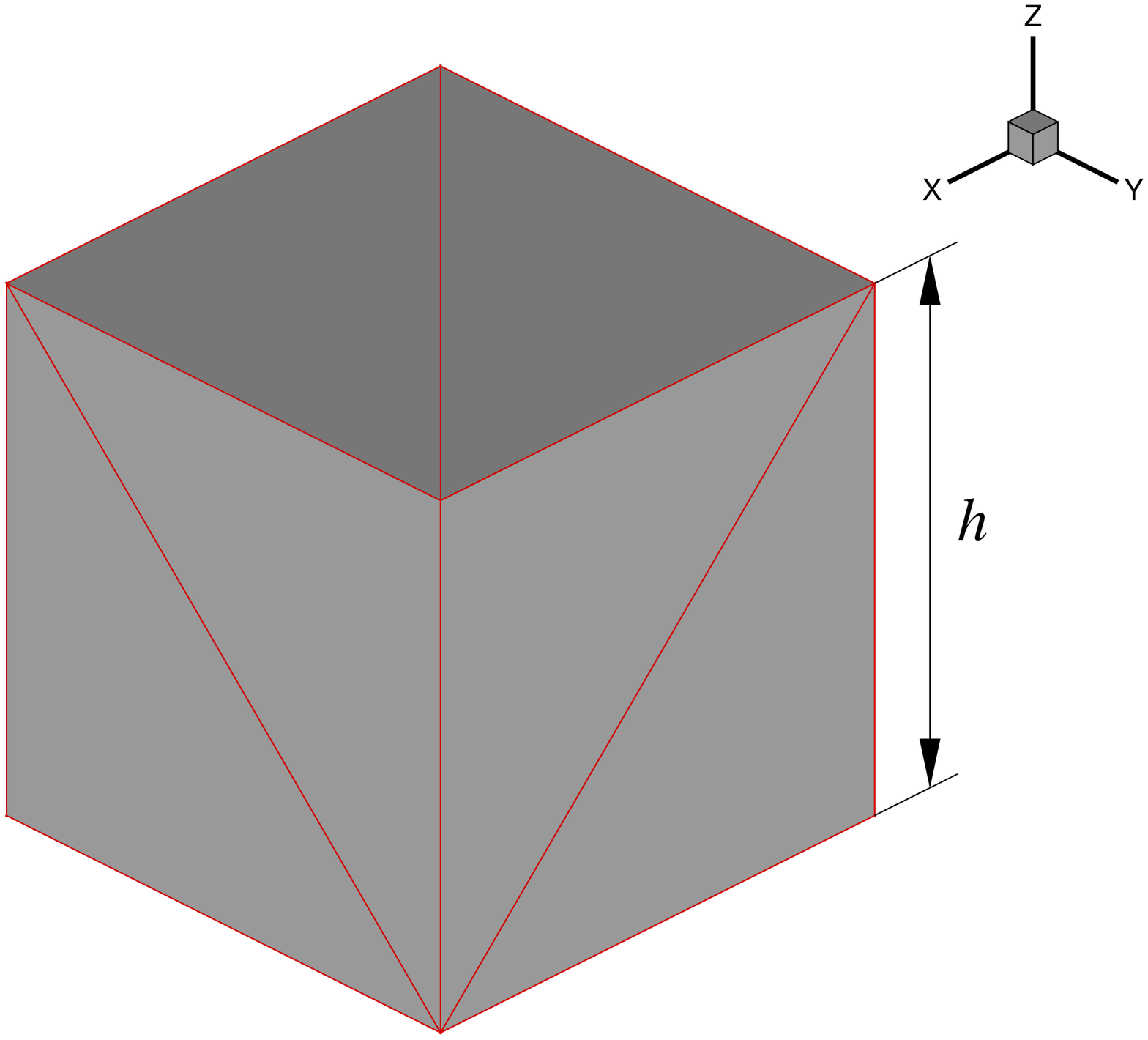}
  \includegraphics[scale=0.27]{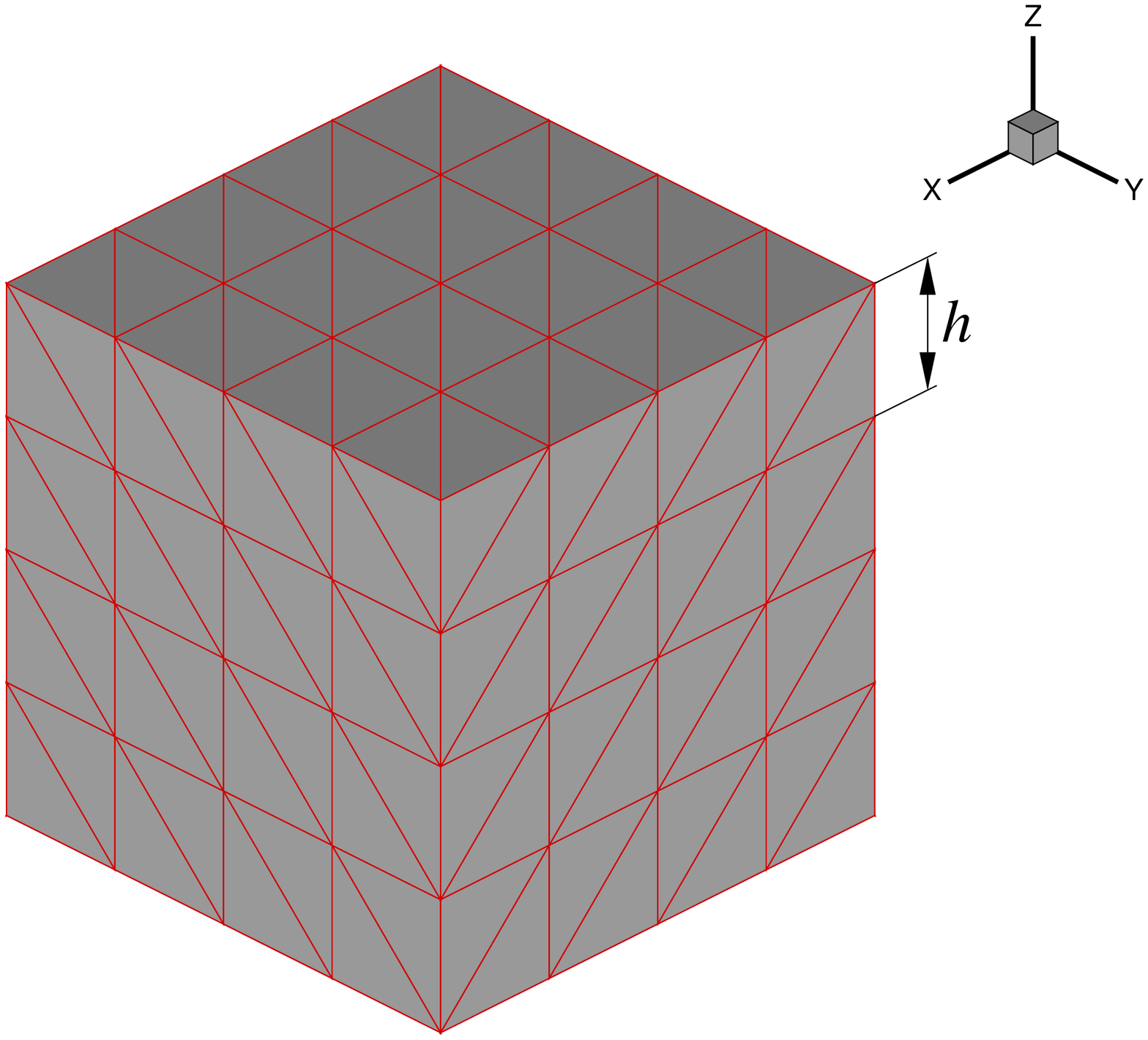}
  \caption{Examples of triangulations used in the convergence study.
    Plots correspond to $h=64 \sqrt{2}\ [\sigma]$ (left) and $h=16
    \sqrt{2}\ [\sigma]$ (right).
    \label{fig:Convergence_triangulation}}
\end{center}
\end{figure}

\begin{figure}
\begin{center}
  \includegraphics[scale=0.5]{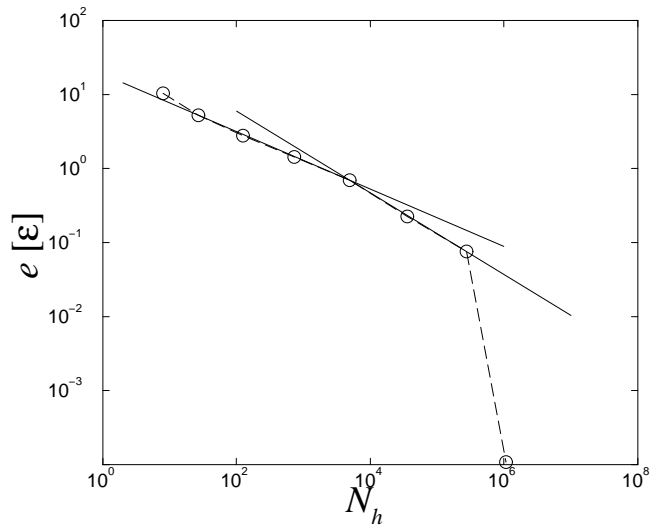}
  \caption{The energy error $e$ as a function of number of
    representative atoms $N_h$ in the sample.
    \label{fig:Convergence_plot}}
\end{center}
\end{figure}

We proceed to assess numerically the rate at which the
quasicontinuum solution converges towards the fully atomistic
solution. The test case is as in the preceding sections. In all
cases, the cluster size $r$ is set to $2 \sqrt{2}\ [\sigma]$ and
the cluster summation weights are computed approximately with a
cutoff parameter $N_c = 2000$. The coarsest mesh is shown in
Fig.~\ref{fig:Convergence_triangulation}a. Subsequent finer meshes
are constructed by regular refinement,
Fig.~\ref{fig:Convergence_triangulation}b, resulting in increasing
numbers of representative atoms $N_h$, with the exception of the
fully resolved case, $h = \sqrt{2}\ [\sigma]$, in which the
representative atoms are placed on all the crystal lattice sites.
In these calculations, the crystal is loaded simply by imparting a
downward displacement $- 0.1 \ [\sigma]$ to the central atom on
the surface labeled $A$ in Fig.~\ref{fig:Convergence_sample}.

Fig.~\ref{fig:Convergence_plot} shows the variation of the energy
error with the number $N_h$ of representative atoms. Three
distinct regimes are evident in this figure. For small values of
$N_h$, the error behaves as $e \sim O({N_h}^{-\alpha})$, with a
convergence rate $\alpha \approx 0.39$. At $N_h=4913$,
corresponding to $h = 4\sqrt{2}\ [\sigma]$, the summation clusters
begin to overlap and the rate of convergence increases to $\alpha
\approx 0.55$. The final drop of the error occurs when full
atomistic resolution is attained.

These results demonstrate the convergence characteristics of the
quasicontinuum method in the extreme case of an applied point
load. When full lattice sums are performed the calculated
convergence rate of $0.55$ approaches the optimal rate of
convergence of linear finite elements applied to linear
elasticity, which is $2/3$. The introduction of cluster summation
rules results in a degradation of the convergence rate. It should
be carefully noted, however, that some of these conclusions may
well depend on the loading geometry. As is well known, the linear
elasticity solution corresponding to a point load diverges under
the point of application of the load and does not possess finite
energy. The finiteness of the energy of atomistic solution owes
entirely to the discreteness of the atomic lattice. In this sense,
the test case considered here is, therefore, a worse case, and it
is possible that the converge of the quasicontinuum method is more
robust for smooth loading.

\subsection{Effect of the remeshing-indicator tolerance}
\label{sec:critical-value-tol}

\begin{figure}
\begin{center}
  \includegraphics[scale=2.0]{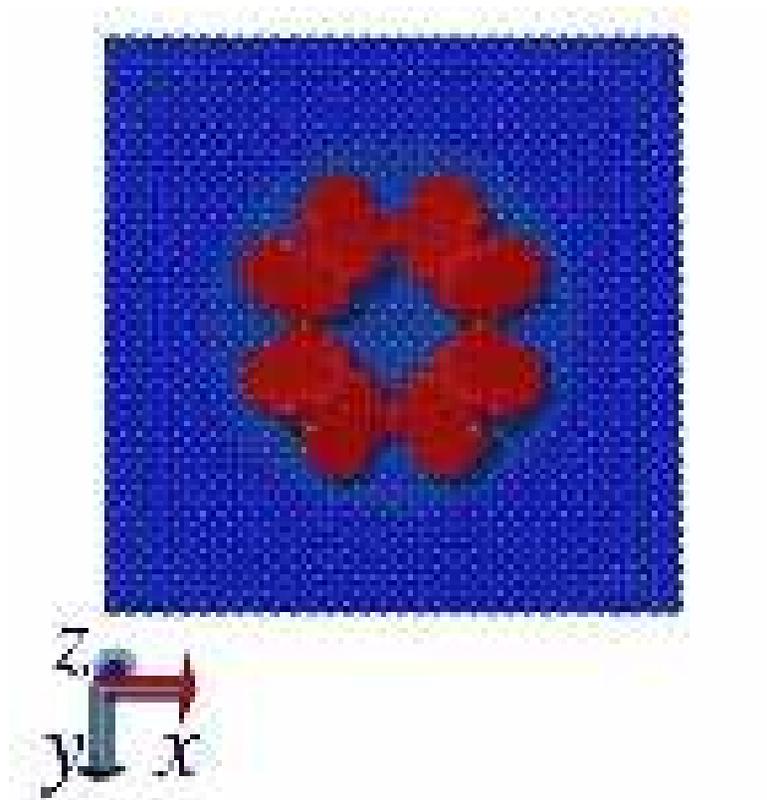}
  \caption{Dislocation structure at the indentor penetration
    $\delta=2.0\ [\sigma]$ predicted by the fully atomistic model.
    The figure displays the energetic atoms (red) underneath the
    crystal surface (blue). \label{fig:Dislocations_atomistic}}
\end{center}
\end{figure}

\begin{figure}
\begin{center}
  \includegraphics[scale=0.5]{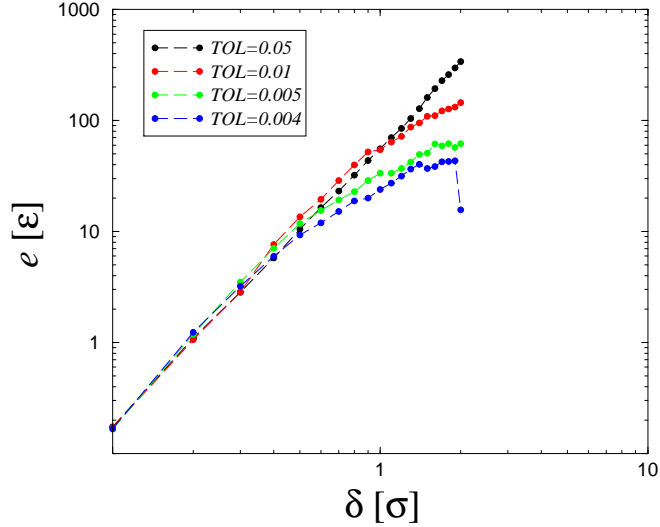}
  \caption{Error $e$ {\it vs} indentation depth
    $\delta$ for different values of the remeshing tolerance
    parameter $TOL$.\label{fig:IndentationError}}
\end{center}
\end{figure}

\begin{figure}[htbp]
  \begin{center}
    \includegraphics[scale=1.5]{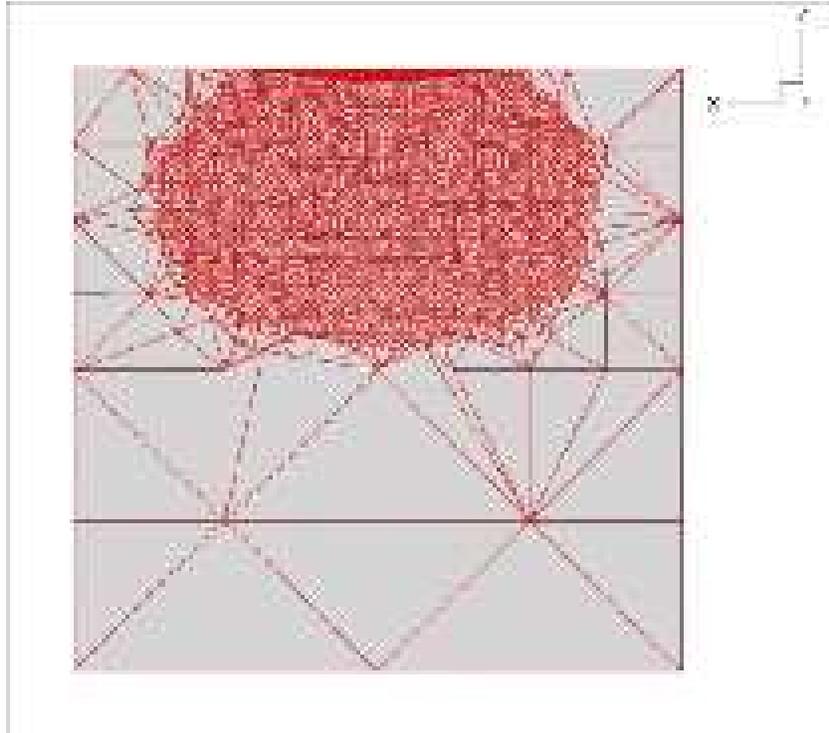}
    \caption{The cross section through the quasicontinuum sample at
    the indentor penetration $\delta = 2.0\ [\delta]$.}
    \label{fig:Final_mesh}
  \end{center}
\end{figure}

\begin{figure}
\begin{center}
  \includegraphics[scale=2.0]{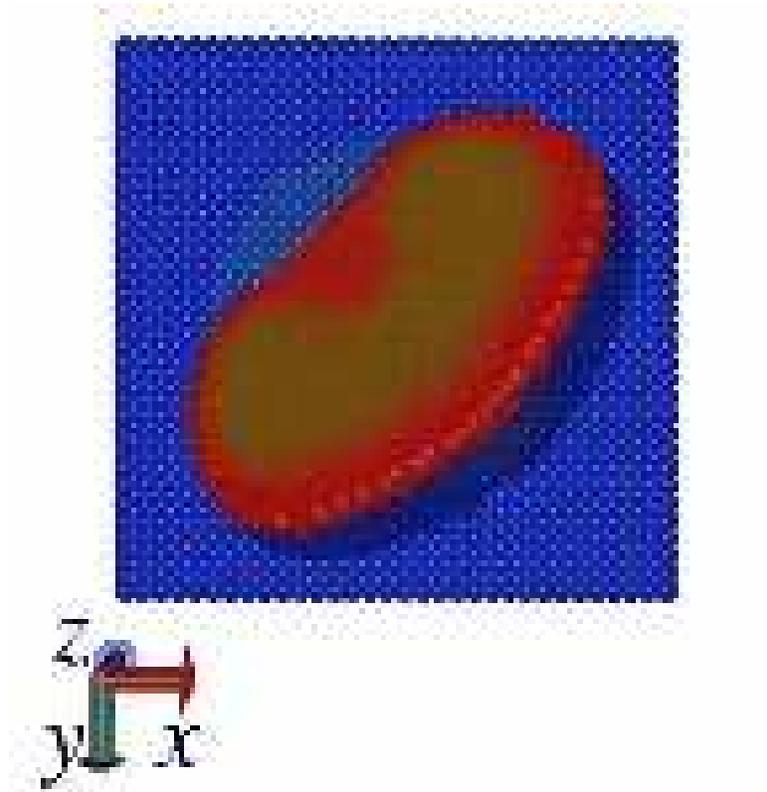}
  \caption{Dislocation structure predicted by the quasicontinuum
  model at indentation depth $\delta = 2.0\ [\sigma]$. The figure
  displays the energetic atoms (red) underneath the crystal surface
  (blue).
    \label{fig:Dislocations_qc}}
\end{center}
\end{figure}

\begin{figure}
  \begin{center}
    \includegraphics[scale=0.5]{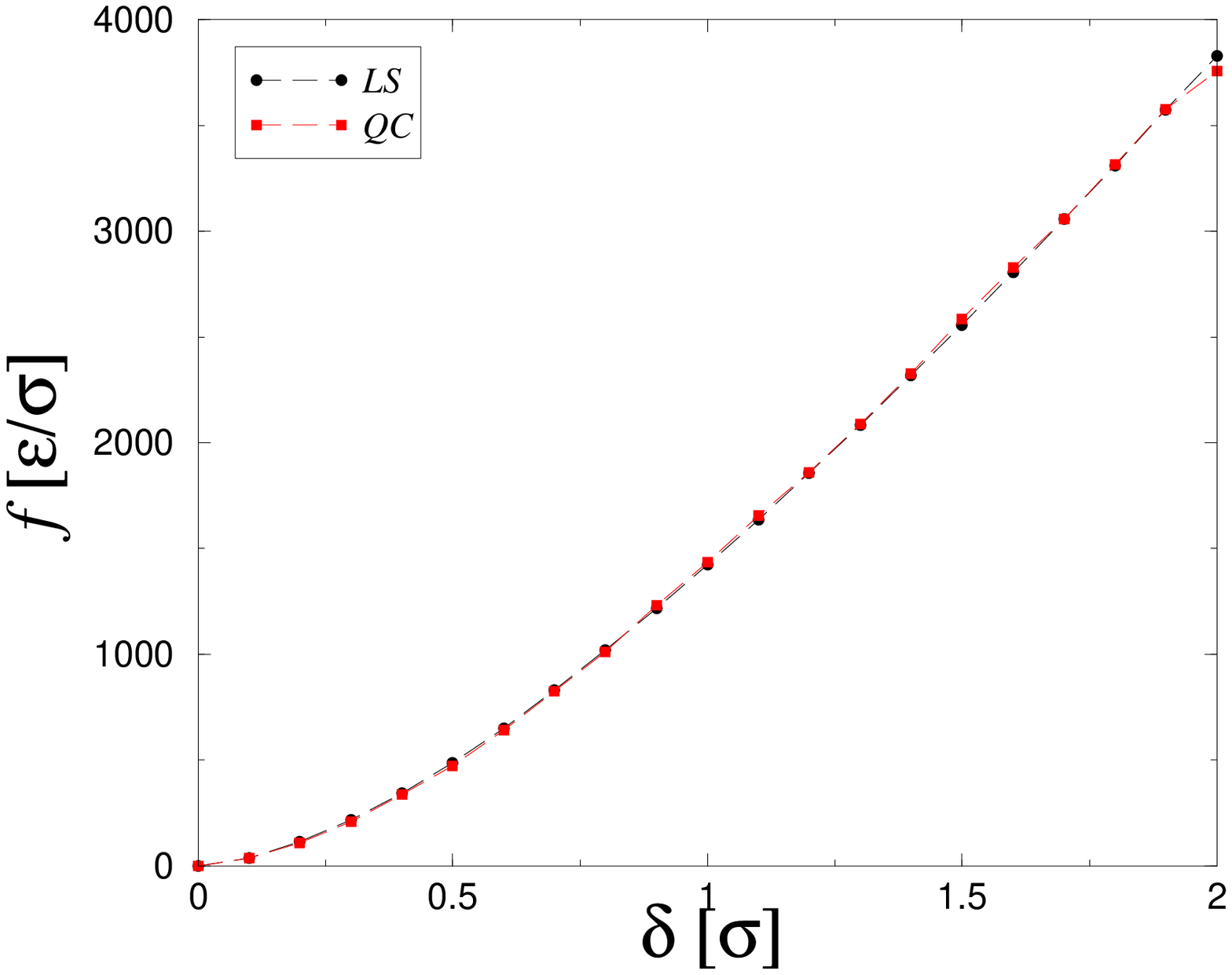}
    \caption{Load {\it vs} displacement curve predicted by
        atomistic (LS) and quasicontinuum (QC) simulations.}
    \label{fig:Force_displacement}
  \end{center}
\end{figure}

The lack of convexity of the energy functional for crystalline
materials allows for lattice defects to develop. Among these
defects, dislocations play a crucial role, as the carriers of the
plastic deformation. One of the appealing features of the
quasicontinuum method is its ability to follow the nucleation and
motion of dislocations inside the crystal. This is achieved by
adaptively providing full atomistic resolution in the highly
deformed regions of the crystal. In the nanoindentation test,
dislocations nucleate under the indentor upon the attainment of a
critical load and subsequently propagate into the crystal.

The insertion of new representative atoms, leading to mesh
refinement, is controlled by criterion (\ref{eq:error_strain2})
and, specifically, by the tolerance $TOL$. A small value of $TOL$
promotes refinement, whereas a large value of $TOL$ inhibits
refinement. Most importantly, an excessively large value of $TOL$
may have the undesirable effect of inhibiting the nucleation of
dislocations under the indentor altogether. On the other hand, an
unduly small value of $TOL$ results in excessive refinement and a
prohibitive computational burden.

In this section we explore this trade-off by way of numerical
testing, with a view to bracketing the optimal value of the
tolerance parameter $TOL$. To this end, we fix the cluster size to
$r=\sqrt{2}\ [\sigma]$, and we use the approximate cluster
summation weights with a cutoff $N_c = 2000$. The indentor is
driven into the crystal at increments of $0.1\ [\sigma]$ up to a
maximum indentation depth of $2.0\ [\sigma]$. The first loading
step does not result in the nucleation of dislocations, even in
the fully atomistic model, and is identical to the calculations
described in the preceding sections. By contrast, at the
indentation depth of $2.0\ [\sigma]$ the presence of dislocation
structures under the indentor is clearly evident in the fully
atomistic model, Fig.~\ref{fig:Dislocations_atomistic}.

The energy error $e$ is plotted in Fig.~\ref{fig:IndentationError}
as a function of indentation depth $\delta$ for different values
of the tolerance parameter $TOL$. Initially, the error is
insensitive to $TOL$, and all curves ostensibly follow the same
linear growth pattern. However, the error curves diverge after a
penetration depth $\delta = 0.5\ [\sigma]$. For large values of
$TOL$, the error continues to grow owing to the inability of the
model to resolve the emerging dislocation structures. Indeed, the
nucleation of dislocations during the test is entirely inhibited
by all but the smallest values of the tolerance, $TOL = 0.004$.
For small values of the tolerance, $TOL = 0.005$ and $0.004$, the
energy error reaches a plateau at roughly $1\%$ of the total
energy, and remains constant or decreases thereafter.

The final mesh at $\delta=2.0\ [\sigma]$ corresponding to a
tolerance of $TOL = 0.004$ is shown in Fig.~\ref{fig:Final_mesh}.
The total number of representative atoms in this configuration is
$N_h = 141,469$, which corresponds approximately to $1/8$ of the
total number $N$ of atoms in the sample. It should be stressed,
however, that since all of new representative atoms are inserted
in the vicinity of the indentor, the ratio $N_h/N$ can be made
arbitrarily small by considering increasingly large samples.

The dislocation structures predicted by the quasicontinuum method
for a tolerance $TOL = 0.004$, Fig.~\ref{fig:Dislocations_qc},
should be contrasted with those predicted by the fully atomistic
model, Fig.~\ref{fig:Dislocations_atomistic}. The figures display
the energetic atoms (red) underneath the surface of the crystal
(blue). Evidently, these structures differ in detail. This is
expected, owing to the massive lack of uniqueness which
characterizes the problem. Thus, as dislocation are nucleated and
the lack of convexity of the energy comes into play, numerous
deformation paths become available to the crystal. Many of these
deformation paths differ only slightly in their energy content and
are, therefore, indistinguishable at the macroscale, e.~g., in
terms of the corresponding force-depth of penetration curves. This
is specially so when the crystal is loaded along a direction of
high symmetry, which results in the possible activation of a large
number of competing slip systems. Under these conditions, a
pointwise comparison of solutions becomes mathematically
meaningless, and the sole meaningful criterion to measure the
quality of a solution is its energy contents. By this criterion,
the quasicontinuum and fully atomistic solutions are of comparable
quality, as their energies differ by less than $1\%$.

This contention is confirmed by Fig.~\ref{fig:Force_displacement},
which compares the force on the indentor $f$ as a function of
penetration depth $\delta$ predicted by the fully atomistic and
quasicontinuum ($TOL = 0.004$) models. Indeed, the two curves are
ostensibly indistinguishable even beyond the onset of dislocation
nucleation. On this basis, in practice choices of the tolerance
$TOL$ in the range of $0.001$--$0.004$ would appear to strike an
adequate balance between accuracy and performance demands.

\section{Summary and Discussion}
\label{sec:Summary}

We have developed a streamlined and fully three-dimensional
version of the quasicontinuum method of Tadmor {\it et al.}
\cite{TadmorOrtizPhillips1996, TadmorPhillipsOrtiz1996} and we
have presented a numerical analysis of its accuracy and
convergence characteristics. As a new addition to the theory, we
have formulated a new class of summation rules in which the
lattice function being summed is sampled over clusters of atoms.
The size of these clusters is an adjustable parameter and controls
the accuracy of the summation rule. We have also presented an
efficient method for computing the requisite summation weights in
linear time. Beyond its usefulness as a numerical scheme for
approximating lattice sums, the cluster approach provides a
systematic means of sampling the behavior of small representative
crystallites, and thus opens a possible avenue for incorporating
additional physics to quasicontinuum models such as diffusion of
solute atoms and thermal lattice vibrations.

We have presented a suite of numerical tests which demonstrate the
accuracy and performance of the method. As expected, the accuracy
of cluster summation rules increases with cluster size.
Furthermore, our numerical tests suggest that the addition of a
single shell of neighbors suffices to stabilize the
rank-deficiency which afflicts node-based summation rules. The
computed convergence rate of the method in problems involving the
application of point loads to crystals is close to linear when the
lattice sums are performed exactly, and decreases somewhat when
the sums are approximated using a cluster summation rule. It is
worth noting that, contrary to continuum problems for which a
well-developed approximation theory exists, a similar
approximation theory for lattice problems appears to be missing at
present, even for linear problems. The development of rigorous
error bounds for finite-element approximations to lattice problems
is a clear worthwhile direction of future research.

Perhaps the most interesting issue among those addressed in this
paper concerns the ability of the quasicontinuum method to
simulate microstructural evolution, which owes largely to mesh
adaptivity. The same lack of convexity which allows for defects
and microstructures to arise in the first place renders solutions
massively nonunique. For a given loading or prescribed
deformation, it is in general possible to find multiple
equilibrium states of a crystal with defects possessing equal or
nearly equal energies. These states are indistinguishable from a
macroscopic point of view, e.~g., in the sense of yielding
identical force-displacement curves. Whether the solution follows
one deformation path or another depends sensitively on small
perturbations of the system, including details of the mesh design.
Under these conditions, the pointwise comparison of solutions is
not particularly meaningful. Our numerical tests suggest that,
with sufficient mesh adaptivity, the quasicontinuum method is
capable of simulating evolving microstructures comparable, in
energetic terms, to those obtained from a full atomistic
calculation.

\section*{Acknowledgments}

The support of the Department of Energy through Caltech's
ASCI/ASAP Center for the Simulation of the Dynamic Behavior of
Solids is gratefully acknowledged.

\bibliographystyle{plain}
{\small%
\bibliography{qc}
}

\end{document}